\documentclass[fleqn,twoside]{article}
\usepackage{espcrc2}

\usepackage{graphicx}
\usepackage[figuresright]{rotating}

\newcommand{\AmS}{{\protect\the\textfont2
  A\kern-.1667em\lower.5ex\hbox{M}\kern-.125emS}}

\title{Production of gauge bosons plus jets in hadronic collisions
        \thanks{Work supported by the European Union under contract 
HPRN-CT-2000-00149.} 
\thanks{Talk given at ICHEP'02 (Amsterdam, 24-31 July 2002).}}
\author{R. Pittau\address{Dipartimento di Fisica Teorica, Universit\`a 
        di Torino and INFN, sezione di Torino, Italy}}
\begin{document}
\newcommand{\bqa}{\begin{eqnarray}}
\newcommand{\eqa}{\end{eqnarray}}
\newcommand{\gev}{\mbox{GeV}}
\newcommand{\lsim}
{\mathrel{\raisebox{-.3em}{$\stackrel{\displaystyle <}{\sim}$}}}
\def\ppbar{p{\buildrel {(-)} \over {p} }}
\begin{abstract}
A computational strategy and a collection of codes are presented
for studying multiparticle final states in hard hadronic collisions.
\vspace{1pc}
\end{abstract}

\maketitle

\section{Introduction}

Multijet final states are characteristic of a large class of important
phenomena present in high-energy hadronic collisions. 
QCD interactions generate multijet final states via 
radiative processes at high orders of perturbation theory 
(pure QCD processes).
Heavy particles in the Standard Model (SM), 
such as $W$ and $Z$ bosons or the top quark, $t$, decay to 
multiquark configurations (eventually leading to jets) 
via electroweak (EW) interactions (mixed EW-QCD processes). 

In addition to the above SM sources, particles possibly present in 
theories beyond the SM
are expected to decay to multiparton final states, and therefore
to lead to multijets. Typical examples are the cascade decays to
quarks and gluons of supersymmetric strongly interacting particles,
such as squarks and gluinos (SUSY-QCD processes).

In addition to fully hadronic multijet final states, a special
interest exists in final states where the jets are accompanied by gauge
bosons. For example
\bqa
\ppbar \to Z +~ jets \to \nu \bar \nu + jets 
\eqa
is an important background to SUSY searches.
Likewise,
\bqa
\ppbar \to W +~ 4~jets \to \ell~\bar \nu_{\ell} + 4~jets 
\eqa
provides the leading source of backgrounds to the identification and
study of top quark pairs in hadronic collisions.

Several parton-level Monte Carlo (MC) event generators exist 
in the literature (such as {\tt VECBOS} \cite{vecbos}, 
{\tt MADGRAPH} \cite{madgraph}, {\tt CompHEP} \cite{comphep},
{\tt GRACE} \cite{grace}, {\tt PAPAGENO} \cite{papageno})
and studies can be performed by assuming that
hard partons can be identified with jets, and that the jets'
momenta are equal to those of the parent partons.  
This simplification gives rough estimates, but
cannot be used in the context of realistic detector simulations, for
which a representation of the full structure of the final state (in
terms of hadrons) is required.  

This full description can be obtained
by merging the partonic final states with shower MC
programs 
where partons are perturbatively evolved through
emission of gluons, and subsequently hadronized.

However, this merging is not always
possible, since common parton-level MC's sum and average over
flavours and colours, and do not usually provide sufficient
information on the flavour and colour content of the events.

In this contribution we review a strategy for the construction of 
event generators for multijet final
states, based on the exact leading-order evaluation of the matrix
elements for assigned flavour and colour configurations, 
and the subsequent shower development and transition into a
fully hadronized final state.

We also present {\tt ALPGEN} \cite{alpgen}, a collection of codes 
realized 
within the framework of the presented approach. 

As a case study, we then focus on
the production of $W Q\bar{Q}+n$ jet final states \cite{wbb} 
(with $Q$ being a massive quark, and $n \le 4$)
and present results for several production 
rates and distributions of interest at the {\tt LHC} and at the 
{\tt TEVATRON}.

\section{The computational strategy}
There are two main sources of problems:
the matrix-element evaluation and the reconstruction of the colour flow.
To illustrate our solutions we take, for simplicity, the case of multigluon
processes, as the extensions to cases with quarks and EW particles
follow the same pattern.  

The process $g g \to 10~g$ has $5 \times ~10^9$ contributing 
Feynman diagrams. To efficiently carry out the computation,  
an algorithm is therefore needed which does not make 
explicit use of Feynman diagrams. 
Fortunately, such an algorithm exists \cite{alpha,kleiss}. 
The key ingredient is the use of a recursive procedure
to determine the matrix element directly from the effective action. 

To better illustrate the problem with the color flow, we take again 
the case of multigluon production.
The scattering amplitude for $n$ gluons with
momenta $p_i^{\mu}$, helicities $\epsilon_i^{\mu}$ and colours $a_i$
(with $i=1,\dots,n$), can be written as
\begin{eqnarray}
  &&\!\!\!\!\!\!\!\!\!\!\!\!M(\{p_i\},\{\epsilon_i\},\{a_i\})=\!\!\!\!\!\!  
   \sum_{P(2,3,\dots,n)}\!\!\!  
\mbox{tr}(\lambda^{a_{i_1}}\,\lambda^{a_{i_2}}\dots
  \lambda^{a_{i_n}}) \; \nonumber \\ 
&& A(\{p_{i_1}\},\{\epsilon_{i_1}\}; \dots 
  \{p_{i_n}\},\{\epsilon_{i_n}\})\,, 
\end{eqnarray}                                          
where the sum extends over all permutations $P_i$ of indices
$(2,3,\dots ,n)$, and the
functions $A(\{P_i\}) $ (known as {\em dual} or {\em
colour-ordered} amplitudes) are gauge invariant,
cyclically-symmetric functions of the gluons'
momenta and helicities. 

Each dual amplitude $A(\{P_i\})$ corresponds to colour flows from one gluon to
the next, according to the ordering specified by the permutation of indices.
Furthermore, each different colour ordering corresponds to a different shower 
evolution of the event, once the hadronization programs are turned on.
Therefore, any realistic event generator should keep track, event by event, 
of the colour flow (and flavour content) to perform a reliable simulation of 
the subsequent evolution into fully hadronized final states.

More details on our algorithm for the colour reconstruction  
can be found in ref. \cite{newapp}.

\section{The {\tt ALPGEN} project}
{\tt ALPGEN} \cite{alpgen} is a collection of codes for the generation of 
multi-parton processes in hadronic collisions implementing 
the computational strategy described in the previous section.

The processes currently available in the package are: 

\begin{tabular}{ll} 
  1)      &  ${p{\buildrel {(-)} \over {p} }} 
           \to  W Q \bar Q + n$~jets ($n \le 4$) \\
  2)      & ${p{\buildrel {(-)} \over {p} }} 
           \to Z/\gamma^\ast Q \bar Q + n$~jets ($n \le 4$) \\
  3)      & ${p{\buildrel {(-)} \over {p} }}  
           \to  W + n$~jets ($n \le 6$) \\
  4)      & ${p{\buildrel {(-)} \over {p} }} 
           \to  Z/\gamma^\ast + n$~jets ($n \le 6$) \\
  5)      & ${p{\buildrel {(-)} \over {p} }}
           \to  n_z~ Z + n_w~ W + n_h~ H  + n$~jets \\
         &
          ($n_z+n_w+n_h+n \le 8$, $n \le 3$) \\
  6)      & ${p{\buildrel {(-)} \over {p} }} 
           \to Q \bar Q + n$~jets ($n \le 6$) \\
  7)      & ${p{\buildrel {(-)} \over {p} }} 
          \to Q \bar Q  Q^\prime \bar Q ^\prime + n$~jets ($n \le 4$)\\
  8)      & ${p{\buildrel {(-)} \over {p} }}
          \to H Q \bar{Q} + n$~jets ($n \le 4$) \\
\end{tabular}

\vspace{0.3cm}

\noindent $Q$ and $Q^\prime$ being massive quarks.
Anywhere, except for processes $5)$, $W$ and $Z/\gamma^\ast$
are considered off-shell, meaning that the corresponding produced 
final state is a pair of leptons.

The documentation and the codes can be found at
{\tt http://mlm.home.cern.ch/mlm/alpgen/}.

\section{Results}
We show a few illustrative examples obtained with {\tt ALPGEN}
for the process ${p{\buildrel {(-)} \over {p} }} \to  W b \bar b + n$~jets.
At the {\tt LHC} one gets the partonic rates of Table~\ref{table1}.
\begin{table*}[htb]
\caption{Partonic rates in pb at the {\tt LHC}, as a function of the total 
number of jets $N_J$ (including $b$ and $\bar b$).}
\label{table1}
\newcommand{\m}{\hphantom{$-$}}
\newcommand{\cc}[1]{\multicolumn{1}{c}{#1}}
\renewcommand{\tabcolsep}{1.5pc} 
\renewcommand{\arraystretch}{1.2} 
\begin{tabular}{l|lllll} \hline
Process      & $N_J=2$ & $N_J=3$ & $N_J=4$ & $N_J=5$ & $N_J=6$
 \\ \hline \hline
 1      & 2.60(1) & 0.63(1) & 0.144(3)   &  0.036(2)  &  0.008(1)\\
 2      &  --     & 2.97(1) & 2.11 (2)   &  1.08(2)   &  0.47(2) \\
 3+4    &  --     & --      & 0.288(1)   &  0.24(1)   &  0.13(2) \\
 5      &  --     & --      &  --        &  0.030(1)  &  0.031(4) \\
 6      &  --     &  --     &  --        &  --        &  0.0010(3) \\
\hline
 Total  & 2.60(1) & 3.60(1) & 2.54(2)    &  1.38(2)   & 0.64(3) \\
\end{tabular}\\[2pt]
The jet defining cuts are $p_T^i\,>\,20~\gev,~ 
\vert \eta_i \vert \,<\, 2.5$ and $\Delta R_{ij}\,>\,0.4$.
\end{table*}
The numbers in the first column refer to the subprocesses given in ref. 
\cite{wbb}. 
The qualitatively new processes appearing when $N_J$
increases are large and are responsible for the growth of the 
partonic rates when going from $N_J= 2$ to $N_J= 3$.

\begin{figure}[htb]
\includegraphics[width=0.47\textwidth,clip]{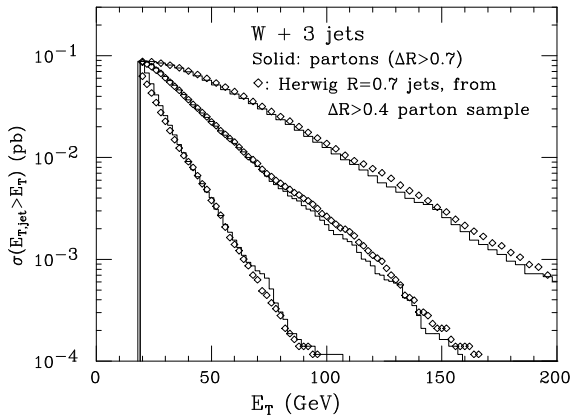}

\vspace{-28pt}

\caption{Inclusive $p_T$ distributions of jets
  at the parton level, with separation cut $\Delta R>0.7$ (solid
  curves), and of fully-showered $R_{jet}=0.7$ jets (dashed curves). 
  These last are
  obtained starting from a full sample of $\Delta R>0.4$ partonic
  events.}
\label{fig1}
\end{figure}

\begin{figure}[htb]
\includegraphics[width=0.47\textwidth,clip]{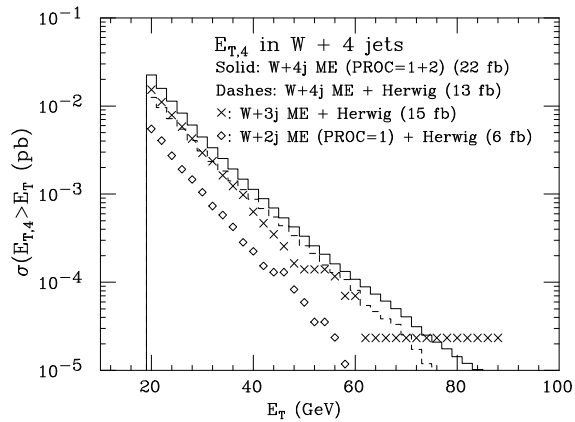}

\vspace{-28pt}

\caption{$p_T$ distributions of jets at the {\tt TEVATRON}:
Matrix Element versus Parton Shower.}
\label{fig2}
\end{figure}

In Figure~\ref{fig1} we show 
a comparison between the $p_T$ distributions of jets 
at the {\tt TEVATRON} before and after the perturbative phase 
of the shower evolution performed by {\tt HERWIG} \cite{herwig}.
A good matching of the jet spectra is
achieved when using generation (parton level) cuts loser than
jet-defining cuts.
Finally, we address the issue of the ability of the shower MC to
correctly predict the rate for hard radiation leading to extra
final-state jets.  
In  Figure~\ref{fig2} we compare the jet rates evaluated with
exact matrix elements with those obtained from hard radiation during the
shower evolution of lower-order parton-level processes.
We find a good agreement in the range $p_T \lsim 45$ GeV, for the 
radiation of one extra jet by {\tt HERWIG}, while
not enough hard radiation is emitted by {\tt HERWIG} to correctly 
predict the emission of 2 extra jets. 

\section{Summary}
The formalism we have outlined
allows us to compute several complex multiparton processes.

{\tt ALPGEN} is a collection of codes realized within the framework 
of the presented approach. It can be used to perform realistic studies 
of multiparticle processes at {\tt TEVATRON} and {\tt LHC}.

{\tt ALPGEN} is ready-to-use and available at
{\tt http//mlm.home.cern.ch/mlm/alpgen/}.

\end{document}